\documentstyle[epsfig,12pt]{article}

\newcommand{\be}{\begin{eqnarray}}
\newcommand{\ee}{\end{eqnarray}}

\begin{document}

\title{{\large {\bf A Diquark Chiral Effective Theory and Exotic Baryons}}}
\author{Deog Ki Hong\thanks{\tt dkhong@pusan.ac.kr},
Young Jin Sohn\thanks{\tt mega@beauty.phys.pusan.ac.kr}
 \\
Department of Physics\\
Pusan National University, Pusan 609-735, Korea \\
\\
Ismail Zahed\thanks{
\tt zahed@zahed.physics.sunysb.edu} \\
Department of Physics and Astronomy\\
State University of New York at Stony Brook, NY 11794, USA\\}
\maketitle


\vspace{0.1in}


\begin{abstract}
A chiral effective theory of quarks and diquarks is formulated and
applied to exotic tetraquarks and pentaquarks. The effective
theory is similar to the chiral quark effective theory with the
addition of diquark degrees of freedom and couplings to quarks.
Chiral symmetry through generalized Goldberger-Treiman relations,
fixes the mass splitting between the even and odd parity diquarks
to be about 600 MeV. We provide an estimate of the parameters of
the effective theory from both the random instanton model and
known data on the low lying scalar nonet. We assess the static
properties of the exotic baryons, such as their masses, magnetic
moments and decay widths. We show that the small decay widths of
the newly reported exotics are largely due to a large tunneling
suppression of a quark between a pair of diquarks. We find
$\Gamma\left(\Theta^+\to K^+n\right)\simeq\,2.5\sim7.0~{\rm MeV}$ and
$\Gamma\left(\Xi^{--}\to\Xi^-\,\pi^-\right)\simeq\,1.7\sim4.8~{\rm MeV}$.
\end{abstract}



\section{Introduction}
A number of recent
experiments~\cite{Nakano:2003qx,Barmin:2003vv,Stepanyan:2003qr}
have reported the occurrence of narrow baryonic excitations with exotic
quantum numbers. Among the members of an anti-decuplet,
an isospin singlet $\Theta^+ (1540)$
and an isospin quadruplet $\Xi_{3/2}(1860)$
were discovered. While several experiments are set to improve on
the current ones, only the masses and upper bounds on the widths of the exotics
were measured. The reported hadronic width of the cascade was smaller than
the experimental resolution of 18 MeV of the NA49 experiment, which
is rather remarkable. The parity of these exotics is expected to
be positive.

Two decades ago the SU(3) version of the Skyrme model predicted
a low lying antidecuplet and a 27-plet with exotic quantum numbers,
in particular an antidecuplet with an isosinglet $\Theta^+$ carrying spin-parity
assignment $1/2^+$ and strangeness +1 with a mass of 1530 MeV and a
width less than 15 MeV~\cite{Diakonov:1997mm}, a remarkable
prediction. This notwithstanding, the occurrence of exotics in QCD
calls for a quark model description.

The observed exotic baryons carry quantum numbers of at least five
quarks. In this spirit, quark-diquark models were recently suggested
~\cite{Jaffe:2003sg,Shuryak:2003zi,Karliner:dt} to account for the
occurrence of low lying multiquark states with exotic quantum numbers.
Scalar and tensor diquarks are strongly bound by color-exchange or
instanton interactions in the color $\overline{\bf 3}$ channel,
providing a natural way for the multiquark states to organize.
The scalars are in the antitriplet of flavor while the tensors
are in the sextet. A pair of P-wave scalars bind to an antistrange
quark to form a low-lying positive parity exotic in the
$\overline{\bf 10}$ of flavor~\cite{Jaffe:2003sg}. A pair of
S-wave scalar and tensor bind to an antistrange quark to form a low
lying positive parity exotic in the ${\bf 27}$ of
flavor~\cite{Shuryak:2003zi}. These exotic arrangements are falsifiable
by experiments~\cite{imai} or by lattice calculations~\cite{Csikor:2003ng}.

In this paper we formulate an effective {\it chiral} theory of quarks
and diquarks with strong correlations in the color antitriplet and
flavor antitriplet (scalar) and sextet (tensor). Since the typical
diquark mass is of order $400-600$ MeV~\cite{ins} it sits mid-way
between the confinement scale with $\Lambda_{QCD}\approx 200$ MeV and the
chiral symmetry breaking scale $\Lambda_{CSB}\approx 4\pi\,f_{\pi}$. The
chiral effective theory we are seeking is analogous to the chiral
quark effective theory formulated by Georgi and Manohar~\cite{Manohar:1983md},
with the addition of diquark degrees of freedom and their couplings to quarks.

\section{The Chiral Effective Lagrangian}

In this section we will discuss some chiral aspects of
the constituent diquarks and derive a chiral Lagrangian
for their interactions with the pseudoscalar nonet of
mesons.

\subsection{Diquarks}

The color antitriplet diquarks are described by local fields. The scalar
diquark is a flavor antitriplet defined as
\begin{equation}
\varphi_{S\alpha}^i(x)=\lim_{y\to x}{
\left|y-x\right|^{\gamma_{S}}\over \kappa_S^2}
\epsilon^{ijk}\epsilon_{\alpha\beta\gamma}
\bar\psi_{cj}^{\beta}(x)\gamma_5\psi_k^{\gamma}(y),
\label{phis}
\end{equation}
and the tensor diquark is a flavor sextet defined as
\begin{equation}
\varphi^{mn}_{T\alpha\,ij}(x)=\lim_{y\to x}
{ \left|y-x\right|^{\gamma_{T}}\over \kappa_T^2}
\epsilon_{\alpha\beta\gamma}\bar\psi_{c\{i}^{\beta}(x)
\sigma^{mn}\psi_{j\}}^{\gamma}(y)\,\,.
\label{phit}
\end{equation}
$\kappa_{S,T}$ are mass scales for the diquark fields,
$\gamma_{S,T}$ are anomalous dimensions of the diquark correlators.
The greek indices denote colors, while $i,j,k=1,2,3$ denote flavors
and $\sigma_{mn}=i[\gamma_m,\gamma_n]/2$ is the covariantized spin
matrix. The tensor diquark is antisymmetric in spin and symmetric in
flavor (the laces in \ref{phit}) . The 3 time components in the tensor
are parity odd, while the 3 space components are parity even.
$\psi_c\equiv C\bar\psi^T$ is a charge conjugated field of a quark field,
$\psi$. The charge conjugation matrix is given as $C=i\gamma_2\gamma_0$.
Being a strong correlator, the diquark field has a decay width
and a form factor, which will be characterized by its couplings to quarks.
The couplings may have momentum dependence.
Note that a chiral transformation turns (\ref{phis})  into a pseudoscalar of
arbitrary flavor. The locality of the fields and the Pauli principle
restricts the pseudoscalar to be again an antitriplet in flavor. Thus,
under a chiral transformation (\ref{phis}) mixes with the pseudoscalar
diquark
\begin{equation}
\varphi_{P\alpha}^i(x)=\lim_{y\to x}{ \left|y-x\right|^{\gamma_{5S}}\over \kappa_{5S}^2}
\epsilon^{ijk}\epsilon_{\alpha\beta\gamma}\bar\psi_{cj}^{\beta}(x)\,\psi_k^{\gamma}(y).
\label{phi5}
\end{equation}
The same arguments apply to the tensor which yields mixing between its
even and odd parity content.

\subsection{Scalars}

The chiral effective Lagrangians for the scalar diquarks require the
pseudoscalars as well. For that we introduce the left/right
combinations of {\it linear} chiral diquark fields

\be
\phi_{R,L}=\frac 12 \left(\phi_S\pm i\phi_P\right)
\label{E1}
\ee
which transform as $(3,1)$ (R) and $(1,3)$ (L)
under rigid $SU(3)_R\times SU(3)_L$. Let $\Sigma=\sigma+i\,\Pi$ be the linear
representation of the scalar plus pseudoscalar nonet that transforms as
$(3,3)$. The chiral symmetric part of the Lagrangian for scalar-pseudoscalar
diquarks to lowest order reads

\begin{eqnarray}
{\cal L}_{3\times 3}=&&
\left|D_\mu\,\phi_R\right|^2
-\Delta^2 |\phi_R|^2 + \frac {g_\pi}{2}
\,\phi_R^\dagger\,\Sigma\,\phi_L \nonumber\\
&&-\frac i{4f_\pi^2} (g_A-1)
\left(D_\mu\phi_L^\dagger \,
(\Sigma^\dagger\partial^\mu\Sigma)\,\phi_L
+{\rm h.c.} \right)\nonumber\\
&&+\left(L\leftrightarrow R\,\,,\Sigma\leftrightarrow
\Sigma^\dagger\right)\,\,. \label{deft0}
\end{eqnarray}
In the vacuum chiral symmetry is spontaneously broken with
$\Sigma$ developing a v.e.v. Specifically, $\Sigma=\xi\,f_\pi\,\xi$
with the non-linear unitary chiral fields $\xi=e^{i\pi/2f_{\pi}}$,
where $\pi=\pi_a\,T_a$ and $T_a$ are $SU(3)$ generators in the
adjoint representation  with a normalization
${\rm tr}\,\left(T_aT_b\right)=1/2\,\delta_{ab}$. The
{\it non-linear} diquark fields follow from (\ref{E1}) through

\be
\varphi_L =\xi\,\phi_L\qquad\qquad \varphi_R=\xi^\dagger\phi_R\,\,.
\label{E2}
\ee
Inserting (\ref{E2}) into (\ref{deft0}) yields the lowest order chiral Lagrangian

\begin{eqnarray}
{\cal L}=&&\left|\left(D_{\mu}+iR_\mu\right)\varphi_R\right|^2
+\frac 12 g_{\pi}\,f_{\pi}\,\varphi_R^\dagger\varphi_L
-\Delta^2\,\varphi_R^\dagger\varphi_R\,+\left(L\leftrightarrow R\right)\nonumber\\
&&+\frac 12 (g_A-1)\left[ \left\{\left(D_{\mu}+iR_\mu\right)
\varphi_R\right\}^\dagger\,A^\mu\,\varphi_R
+{\rm h.c.}\right]+\left(L\leftrightarrow R\right)\nonumber\\
&&-\left(g_S\,\varphi_{Si}^{\alpha}\,
\epsilon^{ijk}\,\epsilon_{\alpha\beta\gamma}\,
\bar\psi_{cj}^{\beta}\,\gamma_5\psi_k^{\gamma}\,+\,g_{P}\,\varphi_{Pi}^{\alpha}\,
\epsilon^{ijk}\,\epsilon_{\alpha\beta\gamma}\,
\bar\psi_{cj}^{\beta}\,\psi_k^{\gamma}+{\rm h.c.}
\right)\nonumber\\
&&+ \,{\cal L}_{\rm int}(\varphi_S, \varphi_P, \psi,
\bar\psi)+{\cal L}_{\chi {\rm Q}}\,\,.
\label{deft}
\end{eqnarray}
All terms retained are chirally invariant for $g_S=g_P$. The
Yukawa couplings $g_S$ and $g_P$ induce mass splitting between
different flavors. The higher-order interactions of diquarks and
quarks are denoted by ${\cal L}_{\rm int}$, which may contain the
(chirally) covariant derivatives and explicit mass breaking
terms.\footnote{Because of chiral symmetry, the pions couple to
quarks and diquarks with a derivative coupling. Hence,  the
pion contribution to the scalar mass, and the magnetic moments of
pentaquarks will be suppressed by $1/\Lambda_{CSB}$, compared to that
of Yukawa coupling of diquarks, as discussed in section 3.}
The usual interactions of quarks and
Nambu-Goldstone bosons are contained in ${\cal L}_{\chi{\rm Q}}$
(chiral quark effective theory).
The relevant degrees of freedom of the diquark chiral effective theory
are constituent quarks, diquarks, gluons, and pions.\footnote{
Having both quarks and diquarks might cause a problem of double-counting.
When the diquark is probed very closely, one sees the structure of diquarks
and eventually diquarks are no longer relevant degrees of freedom.
But, this occurs only when the probing energy-scale is
much higher than the relevant scale for the diquark effective theory.}
When the diquarks are
absent or infinitely heavy, the effective Lagrangian should reduce to
the chiral quark effective theory of Georgi and Manohar. The power-counting
rule should be same and the diquark field scales like $f_{\pi}$.

Since the diquark transforms like
a color antitriplet, the $SU(3)_c$ covariant derivative is given
as

\begin{equation}
D_{\mu}\,\varphi=\partial_{\mu}\,\varphi+ig_sA_{\mu}^a\, {T^a}^*\,\varphi.
\label{covc}
\end{equation}
Chiral symmetry is enforced through

\begin{equation}
iR_\mu =\left( V_\mu-iA_\mu\right)\,,\quad
iL_\mu =\left( V_\mu+iA_\mu\right)\,\,,
\end{equation}
with
\begin{eqnarray}
V_{\mu}={1\over2}\left(\xi^{\dagger}\partial_{\mu}\xi \,+\,
\xi\partial_{\mu}\xi^{\dagger}\right)\,,\quad
 A_{\mu}={1\over2}i\left(\xi^{\dagger}\partial_{\mu}\xi \,-\,
\xi\partial_{\mu}\xi^{\dagger}\right)\,\,.
\label{covf}
\end{eqnarray}

The chiral Lagrangian (\ref{deft}) mixes scalar
and pseudoscalar diquarks through the axial vector current.
The pseudoscalar diquarks are heavy and broad (above the
two quark continuum).
Indeed, the chiral invariant masses in (\ref{deft}) yield
the canonical contributions

\be
{\cal L}_{\rm mass} = -\frac 12 {M_S^2} \,\left|\varphi_S\right|^2
-\frac 12 {M_P^2} \,\left|\varphi_P\right|^2
\label{mass1}
\ee
with

\be
M_S^2=
\Delta^2-\frac{g_\pi f_\pi}{2}\,,\quad
M_P^2=
\Delta^2+\frac{g_\pi f_\pi}{2}\,\,. \label{mass2} \ee The
spontaneous breaking of chiral symmetry pushes the scalar down
(light) and the pseudoscalar up (heavy). The splitting in the
chiral limit is given by \be M_P^2-M_S^2=g_\pi\,f_\pi\,\,,
\label{mass3} \ee which is a generalized Goldberger-Treiman
relation for diquarks. The dimensional parameter $g_\pi$ relates
to the pion-diquark coupling as is evident in the linear
representation. Indeed, $g_{\pi D}=g_\pi/(M_P+M_S)$ is just the
pion-diquark coupling following from (\ref{deft0}) through a
non-relativistic reduction~\cite{FOOT3}. Thus $M_P-M_S=g_{\pi
D}f_\pi$. The pion-diquark coupling is about twice the pion-quark
coupling $g_\pi Q$, and about two-third the pion-nucleon coupling
$g_{\pi N}$. Thus

\be M_P-M_S=g_{\pi D}f_\pi\approx  \frac 23\,g_{\pi N}f_\pi\approx
\frac 23 {M_N}\,\,, \ee where the standard Goldberger-Treiman
relation in the chiral limit was used. In the chiral limit, the
pseudoscalar diquark is about 600 MeV heavier than the scalar
diquark and decouples. The scalar mass is not fixed by
chiral symmetry and will be determined empirically below.

\subsection{Tensors}

The chiral effective Lagrangian for tensor diquarks can also
be constructed in a similar way. For that we separate the
covariant tensor field into its ``electric'' and ``magnetic''
spin components

\be
\phi_E^k =i\phi^{0k}_T\qquad\qquad
\phi_M^k = \frac 12 \epsilon^{krs}\,\phi^{rs}_T
\label{EM1}
\ee
with the color and flavor indices unchanged. The chiral components
of the tensor field follows in the form

\be
\phi_R^k =\frac 12 \left(\phi_M^k-i\phi_E^k\right)\qquad\qquad
\phi_L^k =\frac 12 \left(\phi_M^k+i\phi_E^k\right)\,\,.
\label{EM2}
\ee
Under rigid $\Lambda_{R,L}$ chiral transformations, (\ref{EM2}) transform
as $\Lambda_{R,L}\,\phi_{R,L}\,\Lambda_{R,L}^T$. The lowest order chiral
Lagrangian involving the scalar-pseudoscalar nonet field
$\Sigma=\sigma+i\Pi$ with the tensor diquarks in the linear
representation reads

\begin{eqnarray}
{\cal L}_{3\times 3}=&&
\left|D_\mu\,\phi_R\right|^2
-\Delta_T^2 |\phi_R|^2 + \frac {g_{T\pi}}{2}
\,\phi_R^\dagger\,\Sigma\,\phi_L \Sigma^T\nonumber\\
&&-\frac i{4f_\pi^2} (g_A-1)
\left(D_\mu\phi_L^\dagger \,
(\Sigma^\dagger\partial^\mu\Sigma)\,\phi_L
+{\rm h.c.} \right)\nonumber\\
&&+\left(L\leftrightarrow R\,\,,\Sigma\leftrightarrow \Sigma^\dagger\right)\,\,,
\label{EM3}
\end{eqnarray}
which is analogous to (\ref{deft0}) except for the pion-tensor
pseudoscalar coupling term (third term).

In the non-linear representation we introduce

\be
\varphi_R=\xi^\dagger\,\phi_R\,\xi^{*}\qquad\qquad
\varphi_L=\xi\,\phi_L\,\xi^T\,\,.
\label{EM4}
\ee
in terms of which the chiral effective Lagrangian with pseudoscalar
mesons and tensor diquarks now read

\begin{eqnarray}
{\cal L}=&&\left|\left(D_{\mu}+iR_\mu\,{\bf .} + {\bf .}\,iR_\mu^T\right)\varphi_R\right|^2
+\frac 12 g_{T\pi}\,f^2_{\pi}\,\varphi_R^\dagger\varphi_L
-\Delta^2\,\varphi_R^\dagger\varphi_R\,+\left(L\leftrightarrow R\right)\nonumber\\
&&+\frac 12 (g_A-1)\left[ \left\{\left(D_{\mu}+iR_\mu\,{\bf .} + {\bf .}\,iR^T_\mu\right)
\varphi_R\right\}^\dagger\,A^\mu\,\varphi_R
+{\rm h.c.}\right]+\left(L\leftrightarrow R\right)\nonumber\\
&&-\left(g_T\,\varphi^{mn}_{T\alpha\,ij}\,
\epsilon_{\alpha\beta\gamma}\,
\bar\psi_{c\{i}^{\beta}\,\sigma^{mn}\,\psi_{j\}}^{\gamma}+{\rm h.c.}\right)\,\,.
\label{deftT}
\ee
The dots stand for the position of $\varphi$ inside the bracket.
The Yukawa coupling $g_T$ is the same for even/odd parity
tensors in the chiral limit. (\ref{deftT}) should be added to (\ref{deft})
as the most general chiral effective lagrangian involving both scalars
and tensors in leading order. The mass terms for the electric and
magnetic tensors are

\be M_M^2=\Delta^2-g_{T\pi}\,f_\pi^2\qquad\qquad
M_E^2=\Delta^2+g_{T\pi}\,f_\pi^2\qquad\qquad\,\,, \label{EM41} \ee
leading to a generalized Goldberger-Treiman relation

\be
M_E^2-M_M^2=g_{T\pi}\,f_{\pi}^2\,\,.
\label{EM5}
\ee
Again $g_{T\pi}\,f_\pi/(M_M+M_E)$ plays the role of the pion-tensor
coupling as is clear from (\ref{EM3}). Assuming the latter to be of the
order of $2/3$ the pion nucleon coupling, we conclude that the splitting
between the electric (parity-odd) and magnetic (parity-even)
diquarks is comparable to the splitting between the pseudoscalar
and scalar diquark and of the order of 600 MeV. The magnetic tensor
diquark with positive parity is {\it lighter} than the electric
tensor diquark with negative parity. The magnetic tensor diquark
was used in a recent analysis of the pentaquark~\cite{Shuryak:2003zi}.

In what follows, we will focus on the chiral effective Lagrangian
involving only the scalar diquarks. The inclusion of the tensors
in our analysis is straightforward.

\section{The static properties of exotic baryons}

In this section we will provide both a theoretical
and an empirical determination of the scalar diquark
parameters in (\ref{deft}) by using: {\bf i.\,} results
from the random instanton model to the QCD vacuum;
{\bf ii.\,} known data on the nonet of scalar mesons.
Both extractions will be shown to be consistent.

\subsection{Random Instanton Model}

Some of the parameters in (\ref{deft}) and (\ref{deftT}) have been
measured in the random instanton model of the QCD vacuum
at a cutoff of the order of the inverse instanton size of 3/fm~\cite{ins}.
For completeness we quote the (2 flavor) scalar and tensor parameters
in the table below where the masses
are in MeV and the couplings are in ${\rm GeV}^2$~\cite{ins}
\bigskip
\begin{equation}
\begin{tabular}{|c|c|c|c|c|}
\hline
$M_S$ & $M_T$ & $G_S$ & $G_T$ \\ \hline
$420 \pm 30$ & $570 \pm 20$ & $0.22 \pm 0.01$ &
$0.13 \pm 0.00$\\ \hline
\end{tabular}
\label{massins}
\end{equation}\vskip 0.2in
\noindent
The couplings $G$'s are defined on shell, e.g.~\cite{ins}
\be
\left<0|\varphi_{S\alpha}^i (0)|\varphi_{S\beta}^j (K)\right> =
\frac{G_S}{\sqrt{3}}\,\delta^{ij}\delta_{\alpha\beta}\,\,,
\label{1}
\ee
with $K^2=M_S^2$. (\ref{1}) is calculable in our effective Lagrangian
approach. Indeed, in the flavor symmetric limit we have
(See Fig.~\ref{dix})
\begin{figure}
\epsfxsize=2in
\centerline{\epsffile{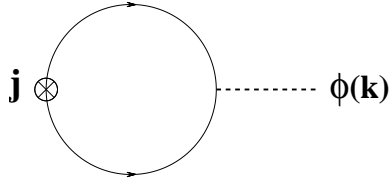}}
%
\caption{Scalar coupling in the effective theory. $\bf j$ is the diquark current.}
 \label{dix}
\end{figure}

\be
\left<0|\varphi_{S\alpha}^i (0)|\varphi_{S\beta}^j (K)\right> =
32\,g_S\,{\bf J} (K) \,\delta_{\alpha\beta}\,\delta^{ij}
\label{2}
\ee
with
\be
{\bf J} (K) =\int\frac{d^4q}{(2\pi)^4}
\frac{m^2-K_+\cdot K_-}{(K_+^2+m^2)(K_-^2+m^2)}\,\,,
\label{3}
\ee
after rotation to Euclidean momenta ($K^2=-M_S^2$).
Here $m$ is the flavor symmetric constituent quark mass, and
$K_\pm=(K/2\pm q)$. Since (\ref{3}) diverges it
requires regularization. For a comparison to the
random instanton model it is perhaps physical to use a
covariant cutoff of the order of the instanton
size used in~\cite{ins}, i.e.
$\Lambda\approx 3/{\rm fm}$. With this in mind,
(\ref{3}) is dominated by a power divergence,
${\bf J}\approx \Lambda^2/(16\pi^2)$. As a result,
the measured on-shell $G_S$ in the random instanton model
translates to our off-shell $g_S$ as
$g_S\approx 2.85\,(G_S/\Lambda^2)$ or $g_S^2\approx 3.03$
at a cutoff scale of the order of $0.6$ GeV. Dimensional
regularization leads to a consistent result for a
comparable renormalization scale. Indeed, in the minimal
subtraction scheme (\ref{3}) reads

\be
{\bf J}^{MS}_E(K) =\left(\frac{M_S}{4\,\pi}\right)^2
\left[1+\frac {3m^2}{4M_S^2} +\frac{m^2}{2M_S^2}\,
{\rm ln}\left(\frac {m^2}{\mu^2}\right)\right]\,\,.
\label{dim}
\ee
For $\mu\approx m\approx M_S$ we get $g_S\approx 1.63\,(G_S/M_S^2)$
or $g_S^2\approx 2.05$. The current results for $g_S$ as extracted
from the random instanton model are consistent with the
empirical estimates we now discuss.


\subsection{Scalar Nonet}

The same effective parameters can also assessed from
experiment. Indeed, the scalar constituent mass of the diquark
can be related to the measured scalar meson masses whereby
the latter are bound diquark-anti-diquark scalars in our
effective theory. From (\ref{deft}) it follows that the
scalar diquark mass reads (see Fig.~\ref{fig1})
\begin{figure}
\epsfxsize=2.5in
\centerline{\epsffile{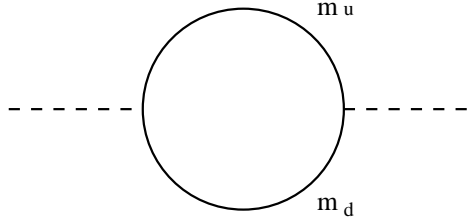}}
%
\caption{One-loop correction to diquark mass.}
 \label{fig1}
\end{figure}
\begin{equation}
M_{jk}^2=M_S^2+{4g_S^2\over \pi^2}\,\left[\left(m_j\,m_k -m_j^2-m_k^2\right)
+{m_j^3\ln\left(m_j^2/\mu^2\right)+m_k^3\ln\left(m_j^2/\mu^2\right)
\over m_j+m_k} \right],
\label{massform}
\end{equation}
where $\mu$ is the renormalization point and
$m_j$ are the constituent quark mass of the $j$-th flavor ($j\ne k$).
The imaginary part of Fig.~\ref{fig1} vanishes below threshold. While
it may broaden the diquarks off-shell, that is in the bound state
configuration of the scalar meson, we expect the broadening to be small and
ignore it.

Assuming the light nonet of scalars to be composed of a bound
diquark and anti-diquark~\cite{Jaffe:1976ig,Black:1998wt}, we
arrive at the mass formula
\begin{eqnarray}
M(a_0)=&M(f_0)&=M_{us}+M_{ds}-B\nonumber\\
M(\kappa)\!\!&=& \!\!M_{us}+M_{ud}-B\label{nonet}\\
M(\sigma)\!\!&=& \!\!2M_{ud}-B\,\,,
\nonumber
\end{eqnarray}
where $B$ is the binding energy of the diquark-anti-diquark pair~\cite{foot1}.
The mass formula (\ref{nonet}) fixes the scalar diquark masses, coupling
and their binding energy in the tetraquark configuration. Thus
\begin{equation}
2M(\kappa)={1\over2}\left[M(a_0)+M(f_0)\right]+M(\sigma),
\end{equation}
which works very well with the experimental values,
$M(a_0)=M(f_0)=980~{\rm MeV}$, $M(\kappa)=800~{\rm MeV}$, and $M(\sigma)
=500\sim 600~{\rm MeV}$.
Taking the renormalization point $\mu=200~{\rm MeV}$ and
$350~{\rm MeV}\le\,M_S\,\le 480~{\rm MeV}$, we get in units of $\rm GeV$
for the masses and the binding energy
\bigskip
\begin{equation}
\begin{tabular}{|c|c|c|c|c|}
\hline
$~\,M_S\,~$ & $M_{us}\,,M_{ds} $ & $~\,M_{ud}\,~$ &
$~~\,B\,~~$ & $~~\,g_S^2\,~~$  \\ \hline
0.35 & 0.56 & 0.38 & 0.15 & 2.64 \\ \hline
0.40 & 0.61 & 0.43 & 0.25 & 2.92 \\ \hline
0.42 & 0.63 & 0.45 & 0.28 & 3.03 \\ \hline
0.45 & 0.66 & 0.48 & 0.34 & 3.19 \\ \hline
0.48 & 0.69 & 0.51 & 0.40 & 3.36 \\ \hline
\end{tabular}
\label{masstable}
\end{equation}\vskip 0.2in
At the renormalization point $\mu=200~{\rm MeV}$ and the scalar mass
$M_S=420\pm30~{\rm MeV}$, obtained in the random instanton model,
the results for the scalar coupling are in good agreement with the
results from the random instanton model.
We note that the scalar couplings for different $M_S$ are small
giving rise to corrections of order $4g_S^2/\pi^2$ or less than
$10\%$. The smallness of the higher-order
corrections indicates that indeed the diquark picture captures
the correct physics of QCD around $400\sim600~{\rm MeV}$. The diquark
anti-diquark bound state should be treated relativistically,
since the binding energy is comparable to the rest mass energy.

In the diquark effective theory, we can also calculate the mass difference
in the anti-decuplet. In particular
\begin{eqnarray}
M(\Xi_{3/2}^{--})-M(\Theta^+)
=(2M_{ds}+m_u)-(2M_{ud}+m_s)=320\sim 180~{\rm MeV},
\label{massrel}
\end{eqnarray}
which works reasonably well with the newly reported splitting from
NA49~\cite{Alt:2003vb}. In reaching (\ref{massrel}) we have used
the fact that the remaining one-gluon exchange between the constituent
diquarks and quarks is flavor blind (perturbative).

\subsection{Magnetic Moments}

Now, we  consider the magnetic moments of exotic baryons, which are
important quantum numbers in photo-production of exotic baryons.
First, we consider $\Theta^+$, whose quark content is $(ud)^2\bar s$.
Being a scalar, the diquark does not carry any magnetic moment. So, in
leading order the magnetic moment of $\Theta^+$ is equal to that of
the anti-strange quark plus that of the orbiting P-wave diquarks,
\begin{equation}
\vec\mu_m(\Theta^+)=\vec\mu_{\bar s}+\vec\mu_L+\delta\vec\mu_m,
\end{equation}
where $\mu_{\bar s}=0.75~{\rm n.m.}$ is the magnetic moment of
the anti-strange quark and $\mu_L$ is the magnetic moment of
the orbiting diquarks, and
$\delta\mu_m$ is a correction due to the quantum fluctuation of
diquarks. Fermi statistics and the range of the effective interaction
force the pair of $(ud)$ diquarks to bind in a P-wave~\cite{Jaffe:2003sg},
resulting in positive-parity pentaquarks. Thus,
for $400~{\rm MeV}\,\le\,M_S\,\le 450~{\rm MeV}$
\begin{equation}
\mu_L=\,{e\over 3}{1\over M_{ud}}\,\simeq\, 1.31\sim 1.46 ~{\rm n.m.}.
\end{equation}
In leading order the radiative correction is shown in Fig.~\ref{fig2}
and reads
\begin{figure}
\epsfxsize=3in
\centerline{\epsffile{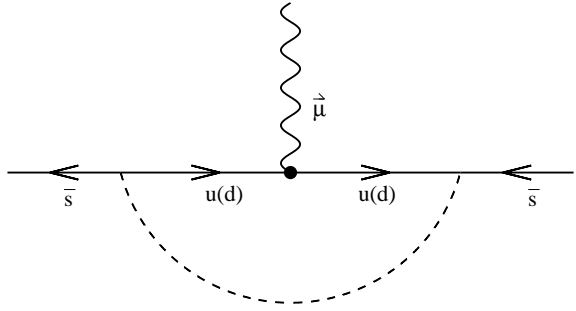}}
%
\caption{One-loop correction to the magnetic moment of anti-strange quark.
The blob denotes the magnetic moments of $u,d$ quarks, $\vec \mu_{u,d}$.}
 \label{fig2}
\end{figure}
\begin{equation}
\delta\mu_m=\mu_u\, {g_S^2\over 16\pi^2}\, {m_u^2\over M_{ds}^2-m_u^2}\,
\left[1-{M_{ds}^2\over M_{ds}^2-m_u^2}\ln\left({M_{ds}^2\over m_u^2}\right) \right]
+\,\left(u\,\leftrightarrow\,d\right)\,.
\end{equation}
Since $g_S^2$ is of order one, the one-loop correction due to
the diquarks to the magnetic moment of the anti-strange quark is quite small,
$\delta\mu_{m}\sim\,5\times10^{-3}~{\rm n.m.}$
We find $\mu_m(\Theta^+)=0.71~{\rm n.m.}$ for $M_S=400~{\rm MeV}$
for $J^P={1\over2}^+$, if
we take $m_u\simeq m_d\simeq 360~{\rm MeV}$ and
$\mu_u=1.98\,{\rm n.m.}$, $\mu_d=-1.10\,{\rm n.m.}$,
while $\mu_m(\Theta^+)=0.56~{\rm n.m.} $ for $M_S=450~{\rm MeV}$.
For $J^P={3\over2}^+$, we get
$\mu_m(\Theta^+)=2.21\sim 2.06~{\rm n.m.} $ with
$400~{\rm MeV}\,\le\,M_S\,\le 450~{\rm MeV}$.
The values we have obtained are in general larger than
the values obtained in the chiral quark-soliton
model~\cite{Kim:2003ay} or those obtained in the QCD sum-rule
estimate~\cite{Huang:2003bu}.
Note that the leading correction is less than a per cent and thus the calculations
are quite reliable.

The present approach shows how to assess the magnetic moments of
the full octet and antidecuplet made of scalar diquarks.
The same approach can be used to assess the magnetic moments of
the exotics in the 27-plet for the case of a scalar and a tensor
pair as discussed above. Also with the chiral effective
Lagrangians (\ref{deft}) and (\ref{deftT}) we may estimate
the masses of the bound exotics and their decay widths using
a 3-body bound state formulation. These issues and others will
be reported elsewhere in a longer analysis. Instead we proceed
to show why generically the newly observed exotics carry small
widths in the diquark description.

\section{Small Decay Widths}
Since the scalar and tensor diquark masses are smaller than the
constituents ($M_{jk}<m_j+m_k$ for $M_{S,T}\,<\,650{\rm  MeV}$),
they are stable against decay near mass shell. In a bound exotic
such as $\Theta^+$ and $\Xi^{--}$ the diquarks orbit in a P-wave.
They are held together with the antistrange quark via colored
Coulomb and confining forces. In such a configuration, the
diquarks are nearby and {\it tunneling} of one of the quarks
between the two diquarks may take place.

Indeed, consider the decay process $\Theta^+\,\to\, K^+\,n$ as
depicted in Fig.~\ref{fig4}. Here a $d$ quark tunnels from a
diquark $ud$ to the other diquark to form a nucleon $udd$ and an
off-shell $u$ quark, which is annihilated by the
anti-strange quark. (If $u$ were to tunnel, the decay is to
$K^0\,p$ with a comparable decay width.)
The decay width is therefore given as
\begin{eqnarray}
\Gamma=\lim_{v\to0}\,\sigma(\bar s +\phi_{ud}+\phi_{ud}\to
K^++n)\,v\,\left|\psi(0)\right|^2,
\end{eqnarray}
where $v$  is the velocity of $\bar s$ in the rest frame of the target
diquark and $\psi$ is the $1S$
wave function of the quark-diquark inside the
pentaquark. The differential cross section for the annihilation
process is then (see Fig.~\ref{fig4})
\begin{eqnarray}
{\rm d}\sigma = {(2\pi)^4\left|\cal M\right|^2\over
4\sqrt{(p_1\cdot p_2)^2-m_s^2M_{ud}^2}}\, 4\,e^{-2S_0}\,{\rm d}\Phi(p_1+p_2;k_1,k_2),
\end{eqnarray}
with the tunneling probability $e^{-2S_0}$ and the phase space
\begin{eqnarray}
{\rm d}\Phi(p_1+p_2;k_1,k_2)=\delta^4(p_1+p_2-k_1-k_2){d^3k_1\over(2\pi)^32E_3}
{d^3k_2\over(2\pi)^32E_4}.
\end{eqnarray}

Assuming that the annihilation amplitude is factorizable (QCD corrections
are small), we find
\begin{figure}
\epsfxsize=2.5in
\centerline{\epsffile{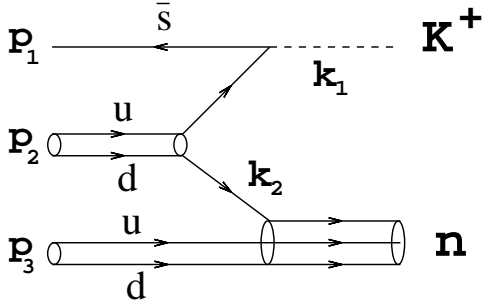}}
%
\caption{Annihilation process $\bar s +\varphi_{ud}+\varphi_{ud}
\to K^++n$. Gluons are suppressed.}
 \label{fig4}
\end{figure}
\begin{eqnarray}
{\cal M}= {g_A\,g\over \sqrt{2}f_K}\,
\bar v_s(p_1)\!\not\! k_1\gamma_5\,{1\over\! \not\!k_1 -\!\not\! p_1-m_u}\,
\gamma_5\, v_d(k_2),
\end{eqnarray}
where $v_{s,d}$ are the wave functions of $\bar s$ and $d_c$, respectively.
Integrating over the phase space and taking $v\to0$, we find the decay width
\begin{equation}
\Gamma_{\Theta^+}\,\simeq\,5.0~ e^{-2S_0}\,
{g^2g_A^2\over 8\pi f_K^2}\,\left|\psi(0)\right|^2.
\end{equation}

Using the WKB approximation, we have
for the tunneling amplitude
\begin{equation}
e^{-S_0}=\left<n\,\right|{T}\,e^{i\int\,{\rm d}^4x\,{\cal L}_{\rm int}}
\left|d\,,\varphi_{ud}\right>\,
\approx e^{-\Delta E \, r_0},
\end{equation}
where $\Delta E\,=\,(m_u+m_d)-M_{ud}$
is the diquark binding energy and $r_0$
is the average distance between two diquarks in the pentaquark (See Fig.~\ref{fig5}).
Of course, the pair in a P-wave senses centrifugation
but this is cutoff at short relative distances by Pauli blocking which
provides a repulsive core, making the height in Fig.~\ref{fig5} likely
higher. So our tunnelling estimate below will be on the lower side.
An estimate of the average distance between two diquarks
in $\Theta^+$, follows from
\begin{equation}
M_{\Theta^+}=2M_{ud}+m_{\bar s}+{2\over M_{ud}\, r_0^2},
\end{equation}
where the third contribution is the rotational energy of diquarks in a
P-wave.
Using the empirical mass $M_{\Theta^+}=1540~{\rm MeV}$ we
find  $r_0=(150~{\rm MeV})^{-1}$ and $\Delta E\,=270~{\rm MeV}$.
The P-wave repulsion estimate
is a bit on the lower side as discussed in~\cite{Shuryak:2003zi}
for a pair of scalar diquarks. This notwithstanding, the WKB
estimate yields a tunnelling amplitude of order
$e^{-1.8}\simeq 0.17$.

The $1S$ wave function of the quark-diquark at the origin can be written as
\begin{equation}
\psi(0)={2\over a_0^{3/2}}{1\over\sqrt{4\pi}},
\end{equation}
where $a_0$ is the Bohr radius of the quark-diquark bound state.
Assuming they are nonrelativistic,  we get by the dimensional analysis
$a_0\simeq\left(2{\overline m}\,B\right)^{-1/2}$,
where ${\overline m}=250~{\rm MeV}$ is the reduced mass and $B$ is the binding
energy of the quark-diquark bound state.
Taking $B=100\sim200~{\rm MeV}$, comparable to the pentaquark binding energy,
$g^2=3.03$ and $g_A=0.75$ from the quark model, we find
\begin{equation}
\Gamma_{\Theta^+}\,\simeq \,2.5\sim 7.0~{\rm MeV}.
\end{equation}

The tunnelling process reduces the decay width by a factor of $50$ to
$100$, compared to normal hadronic decays which do not have any tunnelling
process~\cite{FOOT2}. We claim that the unusual narrowness of the exotic baryons
is naturally explained in the diquark picture. For the $\Xi^{--}_{3/2}$ isospin
quadruplet, we find the height of the potential barrier,
$\Delta E_1\simeq\,270~{\rm MeV}$ with an  average separation of
P-wave strange-diquarks $r_1\simeq(270~{\rm MeV})^{-1}$.
Then, we find
the decay width for $\Xi^{--}_{3/2}\to \Xi^-\,\pi^-$ to be
for $B=100\sim200~{\rm MeV}$
\begin{equation}
\Gamma_{\Xi^{--}_{3/2}}\simeq\,0.51~e^{-2\Delta E_1\,r_1}
\cdot {g^2g_A^2\over 8\pi f_{\pi}^2}\,\left|\psi_u(0)\right|^2
\,\simeq\,1.7\sim4.8~{\rm MeV},
\end{equation}
where $\psi_u$ is the wave function of the anti-up quark and one of the diquarks
in $\Xi^{--}_{3/2}$.
The relative (partial) decay width of $\Xi^{--}_{3/2}$ and $\Theta^+$,
already observed at CERN SPS and at LEPS SPring-8, respectively, is
\begin{eqnarray}
{\Gamma_{\Xi^{--}_{3/2}}\over \Gamma_{\Theta^+}}\,\simeq\,
0.1\, \left({f_K\,\left|\psi_u(0)\right|
\over f_{\pi}\left|\psi_s(0)\right|}\right)^2\,e^{2(\Delta E\,r_0-\Delta E_1\,r_1)}
\,\simeq\, 0.7\,.
\end{eqnarray}
\begin{figure}
\epsfxsize=3.5in
\centerline{\epsffile{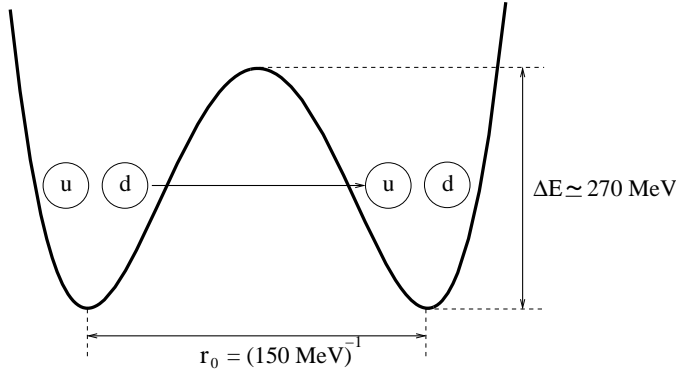}}
%
\caption{Tunnelling of a quark from one diquark to another.}
 \label{fig5}
\end{figure}

\section{Conclusions}
We have formulated a chiral quark-diquark effective theory to analyze
the properties of the newly discovered multiquark exotics. The chiral
effective theory involves the low lying scalar and tensor diquarks
suggested by the random instanton model of the QCD vacuum, and its
parameters are well constrained by the observed scalar mesons treated
as tetraquarks or bound pairs of diquark-anti-diquark. Chiral symmetry
alone shows that the even parity scalar and tensor diquarks are about
600 MeV lighter than their odd parity chiral partners. This maybe the
expected splitting between even-odd parity multiquark exotics with
diquark content.

The diquark chiral effective theory captures correctly the physics of QCD
between the confinement scale and the chiral symmetry breaking scale. It is
similar to the chiral quark effective theory suggested by Georgi and Manohar
with the addition of diquark degrees of freedom and their couplings to
quarks and other particles like pions and gluons.

We have suggested that the smallness of the strong decay widths of
the newly reported exotics is naturally explained as a quark
tunnelling from one pair of diquark to the other. This phenomenon
does not occur in non-exotic strong baryon decays. Our WKB estimate
shows that the strong decay rate in exotics is about two orders
of magnitude suppressed in comparison to non-exotics. We find the
decay width of $\Theta^+\to K^+\,n$ to be
$\Gamma_{\Theta^+}\,=\,2.5\sim7.0~{\rm MeV}$
and the decay width of $\Xi^{--}\to\pi^-\,\Xi^-$,
$\Gamma_{\Xi^{--}}\,=\,1.7\sim4.8~{\rm MeV}$,
unusually narrow as seen by several experiments.

The chiral diquark effective theory proposed here, maybe used to
refine the magnetic moment calculations in the exotic antidecuplet
(scalar-scalar) or 27-plet (scalar-tensor) channels. It can be used
to evaluate the axial charge couplings, and exotic colored Coulomb
bound states made of pairs of diquarks for tetraquark and pentaquark
states. These issues and others will be addressed elsewhere.

\newpage

\section*{Acknowledgements}
\noindent

One of us (DKH) is grateful to T. Kunihiro
for inviting him to the YITP multiquark Workshop  (YITP-W-03-21) of Feb. 2004,
where part of this work was done. He also thanks R. Jaffe for interesting discussions.
The work of DKH and YJS was supported by KOSEF grant number R01-1999-000-00017-0.
The work of IZ was partially supported by the US DOE grant DE-FG-88ER40388.

\bigskip

\section*{Note Added}

\noindent After submitting our paper we became aware of the newly
reported charmed pentaquark by the H1 collaboration in
hep-ex/0403017. In our chiral diquark description with diquark
doublers, the observed charmed pentaquark maybe interpreted as the
$1/2^-$ parity partner of the expected $1/2^+$. Specifically,
$\Theta_s^+$ at 1.5 GeV assumed $1/2^+$ with scalar-scalar
diquarks, would have a chiral partner $1/2^-$ at 2.1 GeV with
scalar-pseudoscalar diquarks. Similarly, its charm counterpart
$\Theta_c^0$ with $1/2^+$ at 2.5 GeV would have a chiral partner
$1/2^-$ at 3.1 GeV. This identification is consistent with a
recent soliton calculation using heavy-spin and chiral symmetry in
hep-ph/0403184.

\bigskip




\vskip 1 in \baselineskip=1.6pt
\begin{thebibliography}{99}
\bibitem{Nakano:2003qx}
T.~Nakano {\it et al.}  [LEPS Collaboration],
Phys.\ Rev.\ Lett.\  {\bf 91} (2003) 012002
[arXiv:hep-ex/0301020].

\bibitem{Barmin:2003vv}
V.~V.~Barmin {\it et al.}  [DIANA Collaboration],
Phys.\ Atom.\ Nucl.\  {\bf 66} (2003) 1715
[Yad.\ Fiz.\  {\bf 66} (2003) 1763]
[arXiv:hep-ex/0304040].


\bibitem{Stepanyan:2003qr}
S.~Stepanyan {\it et al.}  [CLAS Collaboration],
Phys.\ Rev.\ Lett.\  {\bf 91} (2003) 252001
[arXiv:hep-ex/0307018].


\bibitem{Diakonov:1997mm}
D.~Diakonov, V.~Petrov and M.~V.~Polyakov,
Z.\ Phys.\ A {\bf 359}, 305 (1997)
[arXiv:hep-ph/9703373]:
%
M.~Praszalowicz,
Phys.\ Lett.\ B {\bf 575}, 234 (2003)
[arXiv:hep-ph/0308114].

\bibitem{Jaffe:2003sg}
R.~L.~Jaffe and F.~Wilczek,
Phys.\ Rev.\ Lett.\  {\bf 91} (2003) 232003
[arXiv:hep-ph/0307341].

\bibitem{Shuryak:2003zi}
E.~Shuryak and I.~Zahed,
arXiv:hep-ph/0310270.

\bibitem{Karliner:dt}
M.~Karliner and H.~J.~Lipkin,
Phys.\ Lett.\ B {\bf 575}, 249 (2003)
[arXiv:hep-ph/0402260].




\bibitem{imai}K. Imai, talk at YITP workshop on multi-quark hadrons,
Feb. 17-19, 2004, Kyoto, Japan.

\bibitem{Csikor:2003ng}
F.~Csikor, Z.~Fodor, S.~D.~Katz and T.~G.~Kovacs,
JHEP {\bf 0311}, 070 (2003)
[arXiv:hep-lat/0309090];
%
S.~Sasaki,
arXiv:hep-lat/0310014;
T.~W.~Chiu and T.~H.~Hsieh,
arXiv:hep-ph/0403020.


\bibitem{FOOT3}
The non-relativistic reduction
amounts to defining $\phi=({e^{iMv\cdot x}}/{\sqrt{2M}})\,\phi'$ with
velocity $v$ and mean mass $M=(M_P+M_S)/2$ in (\ref{deft0}).


\bibitem{ins}
T.~Schafer, E. Shuryak and J. Verbaarschot,
Nucl. Phys. {\bf B412} (1994) 143.


\bibitem{Manohar:1983md}
A.~Manohar and H.~Georgi,
Nucl.\ Phys.\ B {\bf 234} (1984) 189.




\bibitem{Jaffe:1976ig}
R.~L.~Jaffe,
Phys.\ Rev.\ D {\bf 15} (1977) 267.

\bibitem{Black:1998wt}
D.~Black, A.~H.~Fariborz, F.~Sannino and J.~Schechter,
Phys.\ Rev.\ D {\bf 59} (1999) 074026
[arXiv:hep-ph/9808415].

\bibitem{foot1}
Here we have assumed that the diquark binding energy is
flavor-independent. This is a good approximation, since the typical binding
energy for diquarks is roughly $270~{\rm MeV}$, which is much
bigger than $3(M_{\Lambda}-M_{\Sigma})/4\simeq \,60~{\rm MeV}$.



\bibitem{Alt:2003vb}
C.~Alt {\it et al.}  [NA49 Collaboration],
Phys.\ Rev.\ Lett.\  {\bf 92}, 042003 (2004)
[arXiv:hep-ex/0310014].




\bibitem{Kim:2003ay}
H.~C.~Kim and M. Praszalowicz
arXiv:hep-ph/0308242.



\bibitem{Huang:2003bu}
P.~Z.~Huang, W.~Z.~Deng, X.~L.~Chen and S.~L.~Zhu,
arXiv:hep-ph/0311108.

\bibitem{FOOT2}
Note that the tunnelling suppression amounts to an
effective reduction of $g_A\rightarrow g_A\,e^{-1.8}$ in the
chiral transition charge discussed in~\cite{Shuryak:2003zi}.
The order of the reduction is consistent with their empirical
observation.



\end{thebibliography}
\end{document}